\documentclass[aps,pre
,showpacs
,twocolumn
,superscriptaddress
]
{revtex4-1}
\pdfoutput=1
\usepackage{latexsym}
\usepackage{amstext}
\usepackage{amsmath,amsthm,amssymb}
\usepackage{array}
\usepackage{graphicx}
\usepackage{fancyhdr}
\usepackage{color}
\usepackage{extarrows}
\usepackage{enumerate}
\usepackage{epstopdf}
\usepackage{bm}
\usepackage[colorlinks,
          linkcolor=black,
          citecolor=black,
            urlcolor=blue
          ]{hyperref}
\usepackage{natbib}

\begin{document}

\title{Work Relations Connecting Nonequilibrium Steady States Without Detailed Balance}

\author{Ying Tang}
\email{jamestang23@gmail.com}
\affiliation{Department of Physics and Astronomy, Shanghai Jiao Tong University, Shanghai 200240, China}
\affiliation{Key Laboratory of Systems Biomedicine Ministry of Education, Shanghai Center for Systems Biomedicine, Shanghai Jiao Tong University, Shanghai 200240,
China}
\author{Ruoshi Yuan}
\affiliation{School of Biomedical Engineering, Shanghai Jiao Tong University, Shanghai 200240, China}
\author{Jianhong Chen}
\affiliation{Department of Mathematics, Pennsylvania State University, University Park, Pennsylvania 16802, USA}
\author{Ping Ao}
\email{aoping@sjtu.edu.cn}
\affiliation{Key Laboratory of Systems Biomedicine Ministry of Education, Shanghai Center for Systems Biomedicine, Shanghai Jiao Tong University, Shanghai 200240,
China}
\affiliation{Department of Physics and Astronomy, Shanghai Jiao Tong University, Shanghai 200240, China}
\date{\today}

\begin{abstract}
Bridging equilibrium and nonequilibrium statistical physics attracts sustained interest. Hallmarks of nonequilibrium systems include a breakdown of detailed balance, and an absence of a priori potential function corresponding to the Boltzmann-Gibbs distribution, without which classical equilibrium thermodynamical quantities could not be defined. Here, we construct dynamically the potential function through decomposing the system into a dissipative part and a conservative part, and develop a nonequilibrium theory by defining thermodynamical quantities based on the potential function. Concepts for equilibrium can thus be naturally extended to nonequilibrium steady state. We elucidate this procedure explicitly in a class of time-dependent linear diffusive systems without mathematical ambiguity. We further obtain the exact work distribution for an arbitrary control parameter, and work equalities connecting nonequilibrium steady states. Our results provide a direct generalization on Jarzynski equality and Crooks fluctuation theorem to systems without detailed balance.
\end{abstract}
\maketitle

\section{Introduction}
\label{section1}
Thermodynamics and statistical physics provide a general framework for study of systems in equilibrium. On the contrary, a lack of general principles leads to difficulty in understanding nonequilibrium processes. Recently, a series of equalities referred as fluctuation theorems \cite{evans1993probability,jarzynski1997nonequilibrium,crooks1999entropy,kurchan1998fluctuation,hummer2001free,chernyak2006path,Chetrite2012Quantum,seifert2012stochastic} were established in a wide class of nonequilibrium systems. Remarkably, work equalities such as Jarzynski equality \cite{jarzynski1997nonequilibrium} and Crooks fluctuation theorem \cite{crooks1999entropy} connect work done in nonequilibrium process to free energy differences between equilibriums with detailed balance \cite{hashitsume1991statistical}. For systems without detailed balance, a priori potential function, or Hamiltonian, corresponding to the equilibrium Boltzmann-Gibbs distribution is typically absent. In such cases, thermodynamical quantities based on the potential function, e.g. free energy, could not be defined.

There have been continuous efforts to bridge equilibrium and NESS, such as the Hatano-Sasa's relation \cite{PhysRevLett.86.3463} based on the energetics \cite{sekimoto1998langevin,seifert2012stochastic} and the fluctuation theorems of entropy production \cite{seifert2005entropy,PhysRevE.81.051133,PhysRevLett.108.170603,PhysRevLett.108.240603,PhysRevLett.104.090601}. In these relations, the breakdown of detailed balance is treated as a source of heat or entropy production, and the potential function is usually defined by $\phi\doteq-\ln\rho$, which requires a prior knowledge on the steady state. Alternatively, we investigate in this paper that whether the thermodynamical theory for equilibrium can be naturally generalized to NESS, so that we can consistently define thermodynamical quantities even without a prior known potential function. If this can be done, can work equalities be directly extended to systems without detailed balance, and what is the effect of non-detailed balance on the work distribution?

We consider the Langevin dynamics with explicitly time-dependent control parameters. The detailed balance condition can be violated due to a probability current at NESS. By decomposing the system to a dissipative part and a conservative part, we dynamically construct the potential function, which also leads to the Boltzmann-Gibbs distribution at steady state. We propose a nonequilibrium theory by defining thermodynamical quantities based on the potential function constructed. Thus, the breakdown of detailed balance does not prevent us to use consistent concepts from equilibrium thermodynamics. We further investigate the process of doing work, and obtain the exact work distribution and generalized work equalities. Based on our decomposition, non-detailed balance corresponds to the conservative part, and its role on work equalities can be classified unequivocally.

This paper is organized as follows. In Sec.~\ref{section2}, we provide the background on decomposing the stochastic dynamics in general. In Sec.~\ref{section3}, we present our decomposition for the time-dependent Ornstein-Uhlenbeck process, give the corresponding Fokker-Planck equation, and define the thermodynamical quantities based on the potential function constructed. We then calculate out the work generating functional, and obtain the work distribution and the generalized work equalities in Sec.~\ref{section4}. In Sec.~\ref{section5}, we give detailed discussions on our results. In Sec.~\ref{section6}, we summarize our work. In the appendix \ref{appendix1}, we provide the procedure to calculate out the work generating functional from the path integral method.

\section{Background: Decomposition for Stochastic Dynamics}
\label{section2}
Before demonstrating our method, we briefly review the decomposition for the nonlinear Langevin dynamics with multiplicative noise: $\dot{\textbf{x}}=\textbf{f}(\textbf{x})+N(\textbf{x})\bm{\zeta}(t)$, where 
$\textbf{x}$ is a $n$-dimensional state vector with $\dot{\textbf{x}}$ as its time derivative. The drift force is $\textbf{f}(\textbf{x})$, and $\bm{\zeta}$ is a $k$-dimensional Gaussian white noise with $\langle\bm{\zeta}(t)\rangle=0$, $\langle\bm{\zeta}(t)\bm{\zeta}^{T}(t^{'})\rangle=\delta(t-t^{'})I_{k}$, where $I_{k}$ is the identity matrix, $\delta(t-t^{'})$ is the Dirac delta function, and $\langle\cdots\rangle$ represents the noise average. The symmetric and positive semidefinite diffusion matrix $D(\textbf{x})$ is defined by $N(\textbf{x})N^{T}(\textbf{x})=2\varepsilon D(\textbf{x})$, where the superscript $T$ denotes the transpose, and $\varepsilon$ playing the role of temperature is set to be unity in this paper. Whether detailed balance condition holds or not, a decomposed dynamics equivalent to the above equation was discovered \cite{ao2004potential,ao2007existence}: $[S(\textbf{x})+A(\textbf{x})]\dot{\textbf{x}}=-\nabla_{\textbf{x}}\phi(\textbf{x})+\hat{N}(\textbf{x})\bm{\zeta}(t)$, where $S(\textbf{x})$ defined by $\hat{N}(\textbf{x})\hat{N}^{T}(\textbf{x})=2\varepsilon S(\textbf{x})$ is a symmetric and positive semidefinite matrix, and $A(\textbf{x})$ is an antisymmetric matrix. The scalar potential function $\phi(\textbf{x})$ constructed corresponds to the electrostatic potential in classical physics \cite{yuan2012beyond}, and leads to the Boltzmann-Gibbs distribution $\rho_{ss}(\textbf{x})\propto\exp[-\phi(\textbf{x})]$ at NESS \cite{ao2008emerging}. This decomposition scheme has also been successfully achieved in the discrete Markov process \cite{Ao2013Dynamical}.

\section{Decomposition for Time-Dependent System}
\label{section3}
In the following, we generalize the above decomposition to explicitly time-dependent systems in order to study nonequilibrium dynamical process driven by the external control. To elucidate this procedure without mathematical ambiguity, we take the multi-dimensional Ornstein-Uhlenbeck process added with control parameters as the model:
\begin{align}
\label{Ornstein-Uhlenbeck}
\dot{\textbf{x}}=-F\textbf{x}+\bm{\mu}(t)+\sqrt{2D}\bm{\zeta}(t),
\end{align}
The $n\times n$ force matrix $F$ and diffusion matrix $D$ are constant matrices, i.e. independent of state $\textbf{x}$ and time $t$. The $n$-dimensional explicitly time-dependent vector $\bm{\mu}(t)$ denotes the external control parameter. This model is free of Ito-Stratonovich's dilemma \cite{gardiner2004handbook}. Avoiding mathematical complication, we restrict our discussion to natural boundary condition.

As a paradigm, Eq.~\eqref{Ornstein-Uhlenbeck} contains typical characteristics of a non-equilibrium process: i). it is normally dissipative, $\nabla\cdot F\textbf{x}\neq 0$; ii). the drift force also has nonzero curl $\nabla\times F\textbf{x}\neq 0$, where the cross product for $n$-dimensional vectors is defined as $\textbf{x}\times\textbf{y}\doteq(x_{i}y_{j}-y_{i}x_{j})_{n\times n}$ \cite{ao2004potential}. Thus, the drift force can not be written directly as the gradient of a potential function: $F\textbf{x}\neq-D\nabla_{\textbf{x}}\phi(\textbf{x})$, corresponding to the breakdown of detailed balance \cite{kwon2005structure}. This system covers standard examples studied both theoretically and experimentally: it can describe a Brownian particle dragged in a harmonic potential \cite{PhysRevLett.89.050601,PhysRevLett.91.110601,PhysRevE.67.046102,PhysRevE.76.050101,PhysRevE.74.021111,PhysRevE.80.011117}, a charged Brownian particle in a uniform magnetic field  \cite{PhysRevE.75.032102,PhysRevE.78.052102,PhysRevE.88.022151}, a driven harmonic oscillator with a thermostat \cite{PhysRevLett.97.140603}, and a heat engine in contact with heat reservoirs \cite{PhysRevLett.99.230602}.


\subsection{Construction on the Potential Function}
We find that Eq.~\eqref{Ornstein-Uhlenbeck} can be decomposed as:
\begin{align}
\label{decomposition1}
(S+A)\dot{\textbf{x}}=-\nabla_{\textbf{x}}\phi(\textbf{x},\bm{\alpha}(t))+\sqrt{2S}\bm{\zeta}(t),
\end{align}
with the time-dependent potential function:
\begin{align}
\label{potential1}
\phi(\textbf{x},\bm{\alpha}(t))&=\frac{1}{2}\textbf{x}^{T}U\textbf{x}-\textbf{f}^{T}(t)U\textbf{x}+g(t),
\end{align}
where $g(t)$ is a smooth function of time $t$, and $\bm{\alpha}$ denotes the set of control parameters including $\textbf{f}$ and $g$. We have introduced a new control parameter $\textbf{f}\doteq F^{-1}\bm{\mu}$, because $\textbf{f}$ is usually manipulated in experiments of dragging a Brownian particle \cite{PhysRevLett.89.050601,PhysRevE.76.050101}. The matrices $U$, $S$, $A$ can be solved as follows. Given matrices $F$ and $D$ in Eq.~\eqref{Ornstein-Uhlenbeck}, we first obtain the anti-symmetric matrix $Q$ by the equation:
$FQ+QF^{T}=FD-DF^{T}$. The matrix $U$ is given by $U=(D+Q)^{-1}F$. Then,  $S=[(D+Q)^{-1}+(D-Q)^{-1}]/2$, $A=[(D+Q)^{-1}-(D-Q)^{-1}]/2$, and $A=0$ is equivalent to $Q=0$. The construction on the matrix $U$ has been rigorously proved for arbitrary matrix $F$ in time-independent linear diffusive system \cite{kwon2005structure,Zia2007Probability}.

To better illustrate our decomposition, we consider the example of a charged Brownian particle in a three-dimensional electromagnetic field $\textbf{B}, \textbf{E}(t)$: $m\ddot{\textbf{x}}=-\gamma\dot{\textbf{x}}+(q/c)\textbf{B}\times\dot{\textbf{x}}-k\textbf{x}+q\textbf{E}(t)+\bm{\zeta}(t)$, where $\gamma$, $m$, $q$, $k$ denote respectively friction constant, mass, charge, and stiffness of a harmonic potential. In zero-mass limit $m\rightarrow 0$, this equation of motion directly reduces to the form of Eq.~\eqref{decomposition1}, where $\textbf{B}\times\dot{\textbf{x}}$ serves as $A\dot{\textbf{x}}$. The presence of Lorentz force $(q/c)\textbf{B}\times\dot{\textbf{x}}$ can induce a circular current, which means that a non-vanishing $A$ indicates a breakdown of detailed balance. The conservation of the Lorentz force also implies that breakdown detailed balance does not dissipate based on our decomposition.

According to Eq.~\eqref{decomposition1}, Eq.~\eqref{Ornstein-Uhlenbeck} can be rewritten as: $\dot{\textbf{x}}=-(D+Q)\nabla_{\textbf{x}}\phi+\sqrt{2D}\bm{\zeta}(t)$. Therefore, the drift force $F\textbf{x}$ contains two parts: $-DU\textbf{x}$, which generates a motion towards the origin if $U$ is positive definite, and $-QU\textbf{x}$, which corresponds to a motion along the constant values of $\textbf{x}^{T}U\textbf{x}$. When the noise is switched off and the control parameter is absent, the potential function $\phi$ is the Lyapunov function for the dynamical system \cite{yuan2012beyond,PhysRevE.87.012708,PhysRevE.87.062109,Ma2014Potential}: $d\phi/dt=-\dot{\textbf{x}}^{T}(S+A)\dot{\textbf{x}}=-\dot{\textbf{x}}^{T}S\dot{\textbf{x}}\leq 0$. Thus, the energy dissipates by the presence of the matrix $S$, and the matrix $A$ does not change the potential energy. As a result, $S$ and $A$ respectively correspond to the dissipative part and the conservative part.

\subsection{Fokker-Planck Equation and the Steady State}
The Fokker-Planck equation for Eq.~\eqref{Ornstein-Uhlenbeck} with the control parameter fixed as $\bm{\alpha}_{0}$ is:
\begin{align}
\label{FPE_1} \partial_{t}\rho(\textbf{x},t)=L(\nabla,\textbf{x},\bm{\alpha}=\bm{\alpha}_{0})\rho(\textbf{x},t), \end{align}
where $L(\nabla,\textbf{x},\bm{\alpha})=\nabla^{T}_{\textbf{x}}[D+Q][\nabla_{\textbf{x}}\phi(\textbf{x},\bm{\alpha})+\nabla_{\textbf{x}}]$. The breakdown of detailed balance is equivalent to $Q\neq 0$ by the analysis of probability current at steady state \cite{PhysRevE.87.062109}, and $U=D^{-1}F$ when $Q=0$ \cite{Kwon2011Nonequilibrium}. Whether $Q=0$ or not, the steady state by solving Eq.~\eqref{FPE_1} obeys the Boltzmann-Gibbs distribution $\rho(\textbf{x},t\rightarrow\infty)\propto\exp[-\phi(\textbf{x},\bm{\alpha}=\bm{\alpha}_{0})]$. This realizes an analogy of the Boltzmann-Gibbs distribution from equilibrium to NESS.

With varying control parameter, the modified Fokker-Planck equation for Eq.~\eqref{Ornstein-Uhlenbeck} is \cite{hummer2001free,ao2008emerging}:
\begin{align}
\label{FPE_2}
\partial_{t}\rho(\textbf{x},t)&=\Big[L(\nabla,\textbf{x},\bm{\alpha})
+\dot{\bm{\alpha}}\frac{\partial\ln\rho_{ss}(\textbf{x},\bm{\alpha}(t))}{\partial\bm{\alpha}}\Big]\rho(\textbf{x},t),
\end{align}
where $\rho_{ss}(\textbf{x},\bm{\alpha}(t))$ is the instaneous steady state distribution of the system at time $t$ under the influence of the control parameter. From Eq.~\eqref{FPE_1}, the instaneous steady state also obeys the Boltzmann-Gibbs distribution:
\begin{align}
\label{distribution_BG}
\rho_{ss}(\textbf{x},\bm{\alpha}(t))&\doteq\frac{1}{Z(\bm{\alpha}(t))}\exp[-\phi(\textbf{x},\bm{\alpha}(t))],
\end{align}
where $Z(\bm{\alpha}(t))=\int d\textbf{x}\exp[-\phi(\textbf{x},\bm{\alpha}(t))]$ is the partition function. The system will relax to a NESS with distribution $\rho=\rho_{ss}$ after the control parameter stops varying.


\subsection{Thermodynamical Quantities}
Within the above framework, we give definitions on thermodynamical quantities. From the partition function, the free energy up to a constant is:
\begin{align}
\label{free energy}
F(t)\doteq-\ln Z(\bm{\alpha}(t))=-\frac{1}{2}\textbf{f}^{T}(t)U\textbf{f}(t)+g(t).
\end{align}
We consider the process of varying control parameters from time $t_{0}$  to time $t_{N}$. Then, we define work as:
\begin{align}
\label{work}
W\doteq\int^{t_{N}}_{t_{0}}dt\,\dot{\bm{\alpha}}\frac{\partial\phi}{\partial \bm{\alpha}}=-\int^{t_{N}}_{t_{0}}dt\,\dot{\textbf{f}}^{T}U\textbf{x}+g(t)\Big|_{t_{0}}^{t_{N}}.
\end{align}
This denotes a change of the total potential with respect to the control parameter, which is called the inclusive work \cite{RevModPhys.83.771,jarzynski2007comparison}. It is identical to the work on dragging a Brownian particle in a harmonic potential \cite{PhysRevLett.89.050601,PhysRevE.67.046102}, and moving a charged Brownian particle in a uniform magnetic field \cite{PhysRevE.75.032102}. When $g(t)$ is a constant, $W=\int^{t_{N}}_{t_{0}}dt\,\dot{\textbf{f}}(\partial\phi/\partial{\textbf{f}})$, which is similar to that in the Jarzynski equality \cite{jarzynski1997nonequilibrium}.


We consider an infinitesimal process via the change of the parameter $\bm{\alpha}$. The work performed on the system is $dW=(\partial\phi/\partial\bm{\alpha})d\bm{\alpha}$, the heat absorbed from the environment is $dq\doteq(\partial\phi/\partial\textbf{x})d\textbf{x}$, and the increasing internal energy corresponds to the change of the potential function $d\phi$. Then, the first law holds: $d\phi=dW+dq$. For systems with detailed balance, e.g. the Langevin equation without the nonconservative force \cite{PhysRevLett.86.3463}, these definitions coincide with the energetics \cite{sekimoto1998langevin,seifert2012stochastic}. Our definitions can be generalized to stochastic dynamics with nonlinear drift and multiplicative noise \cite{ao2008emerging}, which are also consistent with the energetics.

\section{Results}
\label{section4}
\subsection{Work Generating Functional}
Next, we study the nonequilibrium process by an arbitrary continuous variation of the control parameter $\bm{\alpha}$. The system is assumed to be at NESS at time $t_{0}$, and relax to another NESS with distribution $\rho_{ss}$ at time $t_{N}$ after the control parameter stops varying. We explicitly calculate the generating functional for the work through the path integral formulation \cite{tang2014summing}:
\begin{align}
\langle e^{-\lambda W}\rangle_{path}&=\int^{\textbf{x}_{N}}_{\textbf{x}_{0}}\mathcal{D}\textbf{x}\exp\{-\int^{t_{N}}_{t_{0}}dt[(\dot{\textbf{x}}+F\textbf{x}-\bm{\mu})^{T}
\notag\\&\quad\times D^{-1}(\dot{\textbf{x}}+F\textbf{x}-\bm{\mu})]/4-\lambda W\},
\end{align}
where $\lambda$ is an introduced parameter, and the measure is given by $\int^{\textbf{x}_{N}}_{\textbf{x}_{0}}\mathcal{D}\textbf{x}\doteq\int^{+\infty}_{-\infty}d\textbf{x}_{N}\cdots\int^{+\infty}_{-\infty}d\textbf{x}_{1}\cdot$ $\int^{+\infty}_{-\infty}\rho_{ss}(\textbf{x}_{0},\bm{\alpha}(t_{0}))$$d\textbf{x}_{0}/|\det(4\pi\tau D)|^{N/2}$ with $\tau$ as the discretized time interval. From the detailed calculation in the appendix, the result appears to be elegant:
\begin{align}
\label{result1}
\langle e^{-\lambda W}\rangle_{path}\notag&=\exp\Big\{\lambda\Delta F|^{t_{N}}_{t_{0}}
-(\lambda-\lambda^{2})\notag\\&\times\int^{t_{N}}_{t_{0}}dt\int_{t_{0}}^{t}dt^{'}\dot{\textbf{f}}^{T}(t)Ue^{-(t-t^{'})F}\dot{\textbf{f}}(t^{'})\Big\}.
\end{align}

\subsection{Work Distribution}
Applying the Fourier's transformation to Eq.~\eqref{result1}, we get the main result about the work distribution:
\begin{align}
\label{work distribution}
P(W)&=\frac{1}{\sqrt{2\pi\sigma^{2}(t)}}\exp\Big[-\frac{(W-\langle W\rangle|_{P(W)})^{2}}{2\sigma^{2}(t)}\Big],
\end{align}
where $\langle\cdots\rangle|_{P(W)}$ denotes the average over work distribution and 
\begin{align}
\label{cumulant}
\langle W\rangle|_{P(W)}&=\Delta F+\frac{\sigma^{2}(t)}{2},
\\\sigma^{2}(t)&=2\int^{t_{N}}_{t_{0}}dt\int_{t_{0}}^{t}dt^{'}\dot{\textbf{f}}^{T}(t)Ue^{-(t-t^{'})F}\dot{\textbf{f}}(t^{'}).
\end{align}

The work distribution solved previously for systems in two or three dimension \cite{PhysRevE.76.050101,PhysRevE.80.011117,PhysRevE.67.046102,PhysRevE.74.021111,PhysRevE.75.032102,PhysRevE.78.052102,PhysRevE.88.022151} serves as a support of our general result. For systems with detailed balance and a known potential function, Eq.~\eqref{work distribution} is consistent with the previous result \cite{speck2005dissipated}. For systems without detailed balance or a prior potential function, Eq.~\eqref{work distribution} has not been obtained. To our knowledge, there is not a single experimental test on Eq.~\eqref{work distribution} for cases without detailed balance. In applications, one need to explicitly calculate out $Q$, $U$, the first and the second cumulant to get full knowledge about the work distribution. Besides, from Eq.~\eqref{free energy} and Eq.~\eqref{cumulant}, we have: $\langle W\rangle|_{P(W)}\geq\Delta F$.

\subsection{Work Equalities}
From Eq.~\eqref{result1}, we derive non-equilibrium work equalities in the following. We emphasis that the results below also depend on the construction of the potential function, and thus they generalize previous work equalities that are typically derived with a prior known potential function \cite{seifert2012stochastic}. Let $\lambda=1$ in Eq.~\eqref{result1}, and we get the generalized Jarzynski equality connecting two NESSs:
\begin{align}
\label{G-Jarzynski}
\langle e^{-W}\rangle_{path}&=e^{-\Delta F}.
\end{align}
The free energy difference is explicitly independent of the magnetic field denoted by the matrix $A$, which is consistent with the Bohr-van Leeuwen theorem of charged particles at equilibrium \cite{van1921problemes,PhysRevE.75.032102,Pradhan2010Nonexistence}.

By definition,
$\langle e^{-\lambda W}\rangle_{F}\doteq\int^{+\infty}_{-\infty}dWe^{-\lambda W}P_{F}(W)$, and $
\langle e^{-(1-\lambda) W^{\dag}}\rangle_{R}\doteq\int^{+\infty}_{-\infty}dW^{\dag}e^{-(1-\lambda)W^{\dag}}P_{R}(W^{\dag})$,
where the subscript $F$, $R$ denote the forward and the reverse process respectively. We consider the time reversed process with $t\rightarrow -t$, and the work $W^{\dag}$ corresponds to the reverse process. From Eq.~\eqref{result1}, we get:
$\langle e^{-\lambda W}\rangle_{F}=\langle e^{-(1-\lambda) W^{\dag}}\rangle_{R} e^{-\Delta F}$, where we consider the state variable with even parity under the time reversal. From these relations, the detailed fluctuation theorem is obtained:
\begin{align}
\frac{P_{F}(W)}{P_{R}(-W)}&=e^{W-\Delta F}.
\end{align}
Besides, as $\langle W\rangle|_{P(W)}$ is odd and $\sigma^{2}(t)$ is even under time reversal, $\langle W^{\dag}\rangle|_{P(W)}\geq\Delta F^{\dag}$ holds for the reverse process.

We can also rewrite Eq.~\eqref{G-Jarzynski} as another form: $\langle\exp\{\int^{t_{N}}_{t_{0}}dt\dot{\bm{\alpha}}[\partial\ln\rho_{ss}(\textbf{x},\bm{\alpha})/\partial\bm{\alpha}]\}\rangle_{path}=1$, which is consistent with the Hatano-Sasa's relation \cite{PhysRevLett.86.3463}. They start from a dynamical equation with a known Hamiltonian perturbed by a nonconservative force in the periodic boundary condition, and their explicit derivation is for the one dimensional case in which $Q=0$. Here, we consider the Langevin dynamics without a prior Hamiltonian, and the potential function governing the dynamics through Eq.~\eqref{decomposition1} is explicitly constructed. We study the natural boundary condition, and non-detailed balance is caused by $Q\neq0$. Then, the role of the detailed balance is classified, which can not be achieved by rewriting the Hatano-Sasa's relation. Furthermore, through the Jensen inequality \cite{gardiner2004handbook}, we have:
$\langle\int^{t_{N}}_{t_{0}}dt\dot{\bm{\alpha}}\{\partial[-\ln\rho_{ss}(\textbf{x},\bm{\alpha})]/\partial\bm{\alpha}\}\rangle_{path}\geq0$,
serving as the second law of thermodynamics.

\section{Discussion}
\label{section5}
Four remarks are in order. First, our result encompasses cases without detailed balance, i.e. $Q\neq 0$. The work equalities are not explicitly dependent of the matrix $Q$. Second, the singularity of the diffusion matrix $D$, i.e. $\det(D)=0$, is pseudo, as $D$ alone does not appear in our result. The appearance of $D$ is in the combination $D+Q$, which is assumed to be non-singular \cite{ao2004potential}. Thus, a natural regulation procedure exists: we can first put perturbative parameters into $D$ to make it non-singular, and safely let these parameters go to zero after the derivation. Third, there are alternative ways to choose the steady state distribution, such as $\rho_{ss}(\textbf{x},\bm{\alpha})=\exp[-\phi(\textbf{x},\bm{\alpha})]/Z(\bm{\alpha}(0))$ \cite{ao2008emerging}. Then, the form of work equalities should be modified correspondingly. Fourth, the explicit form of our result depends on the choice of the control parameter, such as $\textbf{f}$ or $\bm{\mu}$. Our derivation allows us to choose any linear transformation of $\bm{\mu}$ by direct variable substitution.

The potential function Eq.~\eqref{potential1} is uniquely constructed up to a reference point \cite{kwon2005structure}. The non-detailed balance part $Q$ does not lead to more freedom on the construction of the potential function, as the matrix $U$ is uniquely solved for given matrices $F$, $D$. Besides, there can be a shift on the energy reference point denoted by $g(t)$, which is called gauge freedom \cite{ao2008emerging,RevModPhys.83.771}. The work equalities are free of this gauge problem, and are invariant once $g(t)$ is specified. This gauge freedom can be determined by the way of measuring work in experiment. For example, we can choose $g(t)=\textbf{f}^{T}U\textbf{f}/2$ so that $\phi(\textbf{x},\bm{\alpha})=(\textbf{x}-\textbf{f})^{T}U(\textbf{x}-\textbf{f})/2$ to model the experiment of moving the minimum position of the harmonic trap \cite{PhysRevE.67.046102,PhysRevLett.89.050601}. However, this potential function gives zero free energy difference, because the gauge chosen makes the reference point evolve with the system. To figure out the free energy change by varying $\textbf{f}$, the reference point should be fixed, i.e. $g(t)$ be a constant.

Our framework adopts an ``inside'' view of the system,
and an inside observer has the only knowledge on the given dynamics. We explicitly construct a potential function by decomposing such a dynamics. The potential function gives information on the steady state by leading to the Boltzmann-Gibbs distribution, and has the dynamical meaning through Eq.~\eqref{decomposition1}. It also serves as the Lyapunov function governing the dynamics when noise is zero \cite{yuan2012beyond,PhysRevE.87.012708,PhysRevE.87.062109,Ma2014Potential}. Compared with our method, an alternative way to define the potential function by $\phi\doteq-\ln\rho$ \cite{PhysRevLett.86.3463,ge2008generalized} requires a prior knowledge on the steady state.

\section{Conclusion}
\label{section6}
We have developed a thermodynamical theory unifying equilibrium and NESS. We have also obtained explicitly the work distribution and generalized non-equilibrium work equalities, which demonstrate that the free energy difference between NESSs is explicitly independent of non-detailed balance. Our framework can be generalized to the Langevin dynamics with nonlinear drift force and multiplicative noise. The fluctuation relations about the heat $q$ in our framework need to be investigated. For the second order Langevin equation, the parity of state variable under time reversal requires care when applying our method. Our result also remains to be tested experimentally in systems without detailed balance.

\begin{acknowledgments}
We thank Xiaomei Zhu for the critical comments. This work is supported in part by the National 973 Project No.~2010CB529200 and by the Natural Science Foundation of China Projects No.~NSFC61073087 and No.~NSFC91029738.
\end{acknowledgments}

\section*{Appendixes}
\appendix
\section{Detailed Calculation on the Work Generating Functional}
\setcounter{section}{1}
\label{appendix1}
In this appendix, we give the main steps of the explicit calculation on the work generating functional $\langle e^{-\lambda W}\rangle$. From the path integral formulation \cite{tang2014summing}, we have:
\begin{align}
\label{generatingA1}
&\langle e^{-\lambda W}\rangle_{path}
\notag\\&=\int^{+\infty}_{-\infty}\int^{+\infty}_{-\infty}e^{-\lambda W}P(\textbf{x}_{N}t_{N}|\textbf{x}_{0}t_{0})\rho_{ss}(\textbf{x}_{0})d\textbf{x}_{0}d\textbf{x}_{N}
\notag\\&=\int^{\textbf{x}_{N}}_{\textbf{x}_{0}}\mathcal{D}\textbf{x}\exp\Big\{-\frac{1}{4}\int^{t_{N}}_{t_{0}}dt(\dot{\textbf{x}}+F\textbf{x})^{T}D^{-1}(\dot{\textbf{x}}+F\textbf{x})
\notag\\&\quad+\int^{t_{N}}_{t_{0}}dt\Big[\textbf{J}^{T}\dot{\textbf{x}}+\textbf{I}^{T}\textbf{x}-\frac{1}{4}\bm{\mu}^{T}D^{-1}\bm{\mu}-\lambda g(t)\Big]\Big\},
\end{align}
where the measure is given by
$\int^{\textbf{x}_{N}}_{\textbf{x}_{0}}\mathcal{D}\textbf{x}\doteq\int^{+\infty}_{-\infty}d\textbf{x}_{N}\dots\int^{+\infty}_{-\infty}\rho_{ss}(\textbf{x}_{0})d\textbf{x}_{0}\frac{1}{|\det(4\pi\tau D)|^{N/2}}$,
and for convenience of following calculation we have introduced $\textbf{J}^{T}\doteq\frac{1}{2}\bm{\mu}^{T}D^{-1}$, $\textbf{I}^{T}\doteq\frac{1}{2}(\bm{\mu}^{T}D^{-1}F+2\lambda\dot{\textbf{f}}^{T}U)$.

It should be emphasized that the stochastic interpretation in the path integral of Eq.~(\ref{generatingA1}) is Ito's, and thus we use the pre-point discretization in the following. The reason for using Ito's here is that there are a class of equivalent forms of the path integral formulation for the Ornstein-Uhlenbeck process \cite{tang2014summing}. Each has a specific stochastic interpretation and a corresponding Jacobian term on the exponent. After integration, the result from any form is the same, and is independent with the stochastic interpretation. Here, we choose the form with Jacobian zero, and Ito's interpretation should be applied.

Next, we calculate out this path integral in its discretized form by recursion, i.e. integrating $\textbf{x}_{0}$, $\textbf{x}_{1}$, $\dots$, $\textbf{x}_{N}$ in order. As the term $\int^{t_{N}}_{t_{0}}dt[-\bm{\mu}^{T}D^{-1}\bm{\mu}/4-\lambda g(t)]$ and the partition function of the initial distribution $Z(t_{0})$ do not depend on the spatial coordinates $\textbf{x}$, we do not include them when doing the integration on $\textbf{x}$, and will add them up later. Integrating $\textbf{x}_{0}$, we have:
\begin{align}
&\int^{+\infty}_{-\infty}\frac{d\textbf{x}_{0}}{|\det(4\pi\tau D)|^{1/2}}\exp\Big\{-\frac{1}{4\tau}[\textbf{x}_{1}-\textbf{x}_{0}+\tau F\textbf{x}_{0}]^{T}
\notag\\&\quad\times D^{-1}[\textbf{x}_{1}-\textbf{x}_{0}+\tau F\textbf{x}_{0}]+\textbf{J}^{T}_{0}(\textbf{x}_{1}-\textbf{x}_{0})+\tau\textbf{I}^{T}_{0}\textbf{x}_{0}\Big\}
\notag\\&\quad\times\exp\Big[-\frac{1}{2}\textbf{x}_{0}^{T}A_{0}\textbf{x}_{0}+\bm{\mu}_{0}^{T}(D-Q)^{-1}\textbf{x}_{0}\Big]
\notag\\&=\frac{1}{|\det(DB_{0})|^{1/2}}\exp\Big\{-\frac{1}{2}\textbf{x}_{1}^{T}A_{1}\textbf{x}_{1}+[(\tau\hat{\textbf{I}}_{0}^{T}-\textbf{J}_{0}^{T})
B_{0}^{-1}\notag\\&\quad\times V^{T}D^{-1}+\textbf{J}_{0}^{T}]\textbf{x}_{1}
+\tau(\tau\hat{\textbf{I}}_{0}^{T}-\textbf{J}_{0}^{T})B_{0}^{-1}(\tau\hat{\textbf{I}}_{0}-\textbf{J}_{0})\Big\}
\end{align}
where $V=1-\tau F$, $\tau\hat{\textbf{I}}^{T}_{0}\doteq\tau\textbf{I}^{T}_{0}+\bm{\mu}_{0}^{T}(D-Q)^{-1}$,
$B_{0}=V^{T}D^{-1}V+2\tau A_{0}$, and
$A_{1}=(D^{-1}-D^{-1}VB_{0}^{-1}V^{T}D^{-1})/2\tau$.

We then integrate $\textbf{x}_{1}$:
and get $\tau\tilde{\textbf{I}}_{1}=D^{-1}VB_{0}^{-1}(-\textbf{J}_{0}+\tau\hat{\textbf{I}}_{0})+\textbf{J}_{0}+\tau\textbf{I}_{1}$. We repeat these procedures, and after integrating $\textbf{x}_{N}$ we have:
\begin{align}
&\langle e^{-\lambda W}\rangle_{path}\notag\\&\propto\exp\Big\{\tau\sum^{N}_{n=0}(-\textbf{J}_{n}^{T}+\tau\tilde{\textbf{I}}_{n}^{T})B_{n}^{-1}(-\textbf{J}_{n}+\tau\tilde{\textbf{I}}_{n})\Big\}
\notag\\&\doteq\exp\Big\{\frac{\tau^{2}}{2}\Big[\sum^{N}_{i,j=0}\frac{\textbf{J}_{i}^{T}\Xi_{ij}\textbf{J}_{j}}{\tau^{2}}-\frac{2\textbf{I}_{i}^{T}\Pi_{ij}\textbf{J}_{j}}{\tau}+\textbf{I}_{i}^{T}\Gamma_{ij}\textbf{I}_{j}\Big]\Big\},
\end{align}
with $B_{N}=2\tau A_{N}$, $C_{n-1}=D^{-1}VB_{n-1}^{-1}=A_{n}VA_{n-1}^{-1}$,
$\textbf{J}_{n}=\frac{1}{2}D^{-1}\bm{\mu}(t_{n})$, $\tau\tilde{\textbf{I}}_{n}=C_{n-1}(-\textbf{J}_{n-1}+\tau\tilde{\textbf{I}}_{n-1})+\textbf{J}_{n-1}+\tau\textbf{I}_{n}$,
$\tau\hat{\textbf{I}}^{T}_{0}=\tau\textbf{I}^{T}_{0}+\bm{\mu}_{0}^{T}(D-Q)^{-1}$, and $\textbf{I}_{n}=\frac{1}{2}[F^{T}D^{-1}\bm{\mu}(t_{n})+2\lambda(D+Q)^{-1}\dot{\bm{\mu}}(t_{n})](n>0)$.

Here, we have introduced $\Xi_{ij}$, $\Pi_{ij}$, and $\Gamma_{ij}$ for the convenience of calculation. The recursion relations for $\Xi_{ij}$, $\Pi_{ij}$, and $\Gamma_{ij}$ for $i<j$ are:
\begin{align}
\Xi_{ii}&=2\tau B_{i}^{-1}+\Gamma_{i+1,i+1}-2C_{i}^{T}\Gamma_{i+1,i+1}+C_{i}^{T}\Gamma_{i+1,i+1}C_{i},
\notag\\\Gamma_{ii}&=2\tau B_{i}^{-1}+C_{i}^{T}\Gamma_{i+1,i+1}C_{i},
\notag\\\Pi_{ii}&=2\tau B_{i}^{-1}-C_{i}^{T}\Gamma_{i+1,i+1}+C_{i}^{T}\Gamma_{i+1,i+1}C_{i},
\notag\\\Xi_{ij}&=-2(I-C^{T}_{i})\cdots C^{T}_{j-1}\Pi_{jj},
\notag\\\Gamma_{ij}&=2C_{i}^{T}\cdots C^{T}_{j-1}\Gamma_{jj},
\notag\\\Pi_{ij}&=C_{i}^{T}\cdots C^{T}_{j-1}\Pi_{jj},
\notag\\\Pi_{ji}&=\Gamma_{jj}^{T}C_{j-1}\cdots(C_{i}-I).
\end{align}

Using $\Gamma_{NN}=\Xi_{NN}=\Pi_{NN}=A_{N}^{-1}$, $B_{i-1}=A_{i-1}V^{-1}A_{i}^{-1}D^{-1}V$, $C_{i}=A_{i+1}VA_{i}^{-1}$, and $C_{i-1}C_{i-2}\cdots C_{j}=A_{i}V^{i-j}A_{j}^{-1}$, we have for $i<j$:
\begin{align}
\Gamma_{ii}&=A_{i}^{-1},
\notag\\\Xi_{ii}&=2\tau D+\tau^{2}FA_{i}^{-1}F^{T},
\notag\\\Pi_{ii}&=\tau A_{i}^{-1}F^{T},
\notag\\\Gamma_{ij}&=2A_{i}^{-1}(V^{j-i})^{T},
\notag\\\Xi_{ij}&=2\tau^{2}FA_{i}^{-1}(V^{j-i})^{T}F^{T}-4\tau^{2}D(V^{j-i-1})^{T}F^{T},
\notag\\\Pi_{ij}&=\tau A_{i}^{-1}(V^{j-i})^{T}F^{T},
\notag\\\Pi_{ji}&=\tau V^{j-i}A_{i}^{-1}F^{T}-2\tau V^{j-i-1}D.
\end{align}


After adding the term $\int^{t_{N}}_{t_{0}}dt[-\bm{\mu}^{T}D^{-1}\bm{\mu}/4-\lambda g(t)]$ on the exponent and multiplying inverse of the partition function for the initial distribution $1/Z_{0}$, we get the generating functional:
\begin{align}
&\langle e^{-\lambda W}\rangle_{path}
\notag\\&=\exp\Big\{-\frac{1}{2}\textbf{f}^{T}(t_{0})U\textbf{f}(t_{0})+\frac{\tau^{2}}{2}\lambda^{2}\sum^{N}_{i,j=0}[\dot{\bm{\mu}}_{i}^{T}(D-Q)^{-1}
\notag\\&\quad\times A_{i}^{-1}(V^{j-i})^{T}(D+Q)^{-1}\dot{\bm{\mu}}_{j}]+\tau\lambda\sum^{N}_{j=0}\bm{\mu}_{0}^{T}(D-Q)^{-1}
\notag\\&\quad\times A_{0}^{-1}(V^{j})^{T}(D+Q)^{-1}\dot{\bm{\mu}}_{j}+\frac{1}{2}\bm{\mu}_{0}^{T}(D-Q)^{-1}A_{0}^{-1}
\notag\\&\quad\times(D+Q)^{-1}\bm{\mu}_{0}\Big\}.
\end{align}

We can write the above formula in the integral form:
\begin{align}
\label{resultA1}
&\langle e^{-\lambda W}\rangle_{path}
\notag\\&=\exp\Big\{\frac{\lambda^{2}}{2}\int^{t_{N}}_{t_{0}}dt\int^{t_{N}}_{t_{0}}dt^{'}\dot{\bm{\mu}}^{T}(t)\Omega(t,t^{'})\dot{\bm{\mu}}(t^{'})
\notag\\&\quad+\lambda\int^{t_{N}}_{t_{0}}dt^{'}\int^{t_{N}}_{t^{'}}dt\dot{\bm{\mu}}^{T}(t)(D-Q)^{-1}e^{-(t-t^{'})F}\bm{\mu}(t^{'})
\notag\\&\quad+\lambda\int^{t_{N}}_{t_{0}}dt^{'}\bm{\mu}_{0}^{T}(D-Q)^{-1}A_{0}^{-1}e^{-(t^{'}-t_{0})F^{T}}(D+Q)^{-1}
\notag\\&\quad\times\dot{\bm{\mu}}(t^{'})-\frac{1}{2}\textbf{f}^{T}U\textbf{f}|_{t_{0}}\Big\},
\end{align}
with $A^{-1}_{0}=U^{-1}$, and
\begin{align}
\Omega(t,t^{'})=\left\{
\begin{aligned}
&(D-Q)^{-1}U^{-1}e^{-(t^{'}-t)F^{T}}(D+Q)^{-1}, t<t^{'},\\
&(D-Q)^{-1}e^{-(t-t^{'})F}U^{-1}(D+Q)^{-1}, t>t^{'}.
\end{aligned}
\right.
\end{align}
After doing integration by parts and replacing $\bm{\mu}$ by $\textbf{f}\doteq F^{-1}\bm{\mu}$, we obtain the work generating functional:
\begin{align}
\label{resultA2}
\langle e^{-\lambda W}\rangle_{path}\notag&=\exp\Big\{\lambda\Delta F|^{t_{N}}_{t_{0}}
-(\lambda-\lambda^{2})\notag\\&\times\int^{t_{N}}_{t_{0}}dt\int_{t_{0}}^{t}dt^{'}\dot{\textbf{f}}^{T}(t)Ue^{-(t-t^{'})F}\dot{\textbf{f}}(t^{'})\Big\}.
\end{align}


\bibliography{bib}

\end{document}